\documentclass[iop]{emulateapj}
\usepackage{url}

\newcommand{\myemail}{Duncan.Galloway@monash.edu}
\newcommand{\xte}{{\it RXTE}}
\newcommand{\eps}{{\rm erg\,s^{-1}}}
\newcommand{\epcs}{{\rm erg\,cm^{-2}\,s^{-1}}}

\newcommand{\srca}{GS~1826$-$24}
\newcommand{\srcb}{KS~1731$-$26}

\slugcomment{Accepted by ApJ}

\shorttitle{Consistency of neutron-star radius measurements}
\shortauthors{Galloway \& Lampe}

\begin{document}

\title{On the consistency of neutron-star radius measurements \\
from thermonuclear bursts}

\author{Duncan K. Galloway\altaffilmark{1,2,3}
\& Nathanael Lampe\altaffilmark{1}}
\affil{ Monash Centre for Astrophysics (MoCA),
Monash University, VIC 3800, Australia}

\email{\myemail}

\altaffiltext{1}{School of Physics, Monash University}
\altaffiltext{2}{also School of Mathematical Sciences, 
  Monash University}
\altaffiltext{3}{ARC Future Fellow}

\begin{abstract}
The radius of neutron stars can in principle be measured via the
normalisation of a blackbody fitted to the X-ray spectrum during
thermonuclear (type-I) X-ray bursts, 
although few previous studies
have addressed the reliability of such measurements.
Here we examine the apparent radius in a homogeneous sample of
long, mixed H/He bursts from the low-mass X-ray binaries GS~1826$-$24 and
KS~1731$-$26. The measured blackbody normalisation (proportional to the
emitting area) in these bursts is
constant over a period of up to 60~s in the burst tail, even though the
flux (blackbody temperature) decreased by a factor of 60--75\%
(30--40\%).
The typical rms variation in the mean normalisation from burst to burst was
3--5\%, although a variation of 17\% was found between bursts observed
from GS~1826$-$24 in two epochs. 
A comparison of the time-resolved spectroscopic measurements during bursts
from the two epochs shows that the normalisation evolves consistently
through the burst rise and peak, but subsequently increases further in the
earlier epoch bursts. 
The elevated normalisation values may arise from a change in the
anisotropy of the burst emission, or alternatively variations in the
spectral correction factor, $f_c$, of order 10\%.
Since burst samples observed from systems other than GS~1826$-$24
are more heterogeneous, we expect that
systematic uncertainties of at least 10\% are likely to
apply generally to measurements of 
neutron-star radii, unless the effects described here can be corrected
for.
\end{abstract}


\section{Introduction}

Interest has been raised in recent years in the prospects of inferring the
neutron-star mass and radius from thermonuclear bursts. Such a possibility can
provide stringent constraints on the neutron-star equation of state, which
remains uncertain \cite[e.g.][]{lp07}.
From combining measurements of the blackbody normalisation (from
time-resolved X-ray spectral fits to the burst spectra), Eddington
luminosity, and the distance, confidence limits on the mass and radius of
several neutron stars in low-mass X-ray binaries (LMXBs) has been
estimated (\citealt{ozel09a,guver10a,guver10b,ozel11a}; see also
\citealt{slb10}). Each of these
quantities is challenging to measure alone, but particularly substantial
systematic errors are known to affect the blackbody normalisation.
The normalisation can rise or fall steadily
throughout the burst tail, depending roughly on the duration of the burst
\cite[]{bhatt10a}.
Furthermore, in some sources different bursts provide significantly
different normalisation values. \cite{guver11a} quantified some of these
effects at the 3--7\% level, although in an earlier analysis of the burst
source EXO~1745$-$248 found apparent radii measured from two bursts
varying by 11\% \cite[]{ozel09a}. 
Additionally, analysis of bursts from another globular cluster source,
4U~1724$-$307, found blackbody normalisations which varied by a factor of
$\approx2$ (i.e. a 40\% variation in the inferred radius) between short-
and long-duration bursts \cite[]{suleimanov11b}.

Independent of the issues for measurement, are uncertainties about
precisely how the neutron star's atmosphere affects the emerging
radiation. Although the burst spectra are typically found observationally
to be consistent with a blackbody
\cite[e.g.][]{swank77,kuul02a}, scattering effects have long been
understood to distort the spectrum sufficiently to bias the measured
temperature \cite[e.g.][]{lth84,lth86}. This distortion is usually parameterised
via a spectral distortion factor $f_c=T_{\rm bb}/T_{\rm eff}$ where
$T_{\rm bb}$ is the measured blackbody (or colour) temperature, and $T_{\rm
eff}$ is the effective temperature of the atmosphere. Most recent
work adopt a narrow range of $f_c=1.3$--1.4 at burst luminosities
well below the Eddington limit \cite[e.g.][]{madej04}, and often
neglect the variations in $f_c$ that may arise during the burst 
\cite[e.g.][]{suleimanov11a}. 
A more rigorous approach involves
fitting the observed variation in the blackbody
normalisation (in response to the changing $f_c$) as a function of flux
\cite[e.g.][]{suleimanov11b}, although 
the predicted model curves cannot yet reproduce the range of observed
behaviour.

Samples of bursts accumulated from individual sources can be extremely
heterogeneous in their properties.
From low-duty cycle observations featuring
gaps due to Earth occultations and other conditions, it is usually
impossible to be confident about the burst recurrence time, in which case
the detailed ignition conditions, fuel composition, and even the
completeness of thermonuclear burning are also uncertain. Under such
conditions, it is difficult to disentangle the various systematic
influences which might influence the normalisation measurements
\cite[e.g.][]{guver11a}.
Here we investigate the intrinsic reproducability of burst normalisation
measurements using a uniform, homogeneous sample of bursts, from the
low-mass X-ray binaries \srca\ and \srcb. We use {\it Rossi X-ray Timing
Explorer}\/ ({\it RXTE}) Proportional Counter Array (PCA) data to
test for intrinsic 
systematic effects which might influence burst normalisation 
measurements beyond any additional effects which might arise from
variations in the ignition conditions and fuel composition.
In a related paper, 
\citealt{zamfir12a}
(submitted to ApJ; hereafter Z11)
used the same data to infer the mass and radius of \srca.

\section{Observations and Analysis}

Few bursting sources exhibit trains of bursts with consistent lightcurves
or recurrence times.
The best known example is
GS~1826$-$24, so far unique for its consistently regular burst behaviour,
and high degree of uniformity between successive burst lightcurves
\cite[e.g.][]{gal03d}.
Comparison of the burst
behaviour and lightcurves suggest that the system accretes mixed helium
and hydrogen at roughly solar mass fraction
\cite[]{heger07b}.
We used observations of \srca\ taken with the Proportional Counter Array
\cite[PCA;][]{xte96} onboard the {\it Rossi X-ray Timing Explorer}\/ ({\it
RXTE}), from the
catalogue of bursts detected over the mission lifetime  (\citealt{bcatalog};
hereafter G08). 
The flux-recurrence time relationship for this sample has been extensively
studied by \cite{thompson08}.
Optical photometry of the mass donor suggests an orbital period of
2.25~hr \cite[]{mrp10}. However, several
alias peaks are present in the periodogram, and it is possible one of
these (particularly at 2.05~hr) represents the true orbital period.

We performed a search of the G08 sample for additional examples of
regular, consistent bursts.
Recurrence time provides the most obvious way to detect regular bursts,
although for instruments in low-Earth orbit like \xte, recurrence time
measurement is confounded by regular interruptions due to
occultations of the star by the Earth.
An alternative approach is to test for consistency of the burst light
curve, via commonly-used parameters measuring the duration.  The ratio
$\tau$ of the fluence $E_b$ to the peak flux, as used by G08, provides a
simple way of comparing light curves. A scatter plot of $\tau$ against
$E_b$ for a source with consistent bursts, will show strong clustering,
and \srcb\ provides the next best example after \srca.
Extensive observations of KS~1731$-$26 by the {\it BeppoSAX}/WFC suggest
that, at times, this system exhibits regular bursts more frequently than 
GS~1826$-$24 \cite[see e.g. Fig. 3 from][]{corn03a}. Despite this, little
attention has been directed at the burst behaviour of this
system. One reason is that, unlike GS~1826$-$24, KS~1731$-$26 exhibits
both radius-expansion bursts, and burst oscillations \cite[]{muno00};
previous analyses have focussed largely on these phenomena.
A series of \xte\/ observations of KS~1731$-$26 were made in 2000 August
and September, detecting a total of 14 bursts with highly consistent
lightcurves. WFC observations also made during this time show that the
burst behaviour was quite regular, with recurrence times of between
2--3~hr.
An additional example, with a lightcurve very similar to those observed in
2000, was detected on 1999 Aug 26.
Soon after the 2000 observations, the source faded to quiescence, and has
not been active since \cite[]{wij01d}.

No other sources
exhibit clustering as tight in the rest of the G08 sample; the next best
examples are 4U~1323$-$62 and 4U~0836$-$429.
4U~1323$-$62 is quite well-known for it's regular burst behaviour (e.g.
G08), and in
the \xte\/ sample, many of the bursts were long with $\tau=28\pm2$~s and
similar fluences, of $E_b=(100\pm10)\times10^{-9}\ {\rm erg\,cm^{-2}}$.
However, several of the bursts were closely followed by weaker events only
a few minutes after the first. This behaviour has been observed in a
number of systems \cite[e.g.][]{keek10}, and is thought to arise from
delayed ignition of material left over from the initial burst. Such events
are likely not suitable for a detailed test of the consistency of
blackbody normalisations between bursts.
During a period of activity in 2003--4, 4U~0836$-$429 exhibited long
($\tau=22\pm4$~s) bursts at roughly similar fluences, distributed
approximately as a Gaussian with $E_b=(270\pm70)\times10^{-9}\ {\rm
erg\,cm^{-2}}$. Despite good coverage of the outburst, the bursts were
apparently not strictly periodic in nature, and there may also be
contributions to the source flux from
a nearby HMXB pulsar which is within the PCA field
of view. Thus, we also exclude this source from our study.

Where not explicitly stated, the data analysis procedures
are as in G08. Time-resolved spectra in the range 2--60~keV covering the
burst duration were extracted on intervals as short as 0.25~s during the
burst rise and peak, with the bin size increasing gradually into the burst
tail to maintain roughly the same signal-to-noise level. A
spectrum taken from a 16-s interval prior to the burst was adopted as the
background.
We re-fit the spectra over the
energy range 2.5--20~keV using the
revised PCA response matrices, v11.7\footnote{see
{\url
http://www.universe.nasa.gov/xrays/programs/rxte/pca/ doc/rmf/pcarmf-11.7}}
and adopted the recommended systematic error of 0.5\%. The fitting was
undertaken using {\sc XSpec} version 12.
In order to accommodate spectral bins with low count rates, we adopted
Churazov weighting.
No correction for instrumental deadtime was applied to the spectra. 

We modelled the effects of interstellar absorption, using a
multiplicative model component ({\tt wabs} in {\sc XSpec}), with the
column density $n_H$ frozen at $4\times10^{21}\ {\rm cm^{-2}}$ (for \srca,
e.g. \citealt{zand99})
and $1.3\times10^{22}\ {\rm cm^{-2}}$ (for \srcb, e.g. \citealt{cackett06}).
In the original analysis carried out by G08,
the
neutral absorption was determined separately for each burst, from the mean
value obtained for spectral fits carried out with the $n_H$ value free to
vary. This has a negligible effect on the fluxes, but can introduce
spurious burst-to-burst variations in the blackbody normalisation.
We also computed averaged lightcurves of blackbody spectral parameters
for subsets of bursts from \srca,
following the procedure adopted by \cite{gal03d}.

\section{Results}
\label{sec1}

We first explored a variety of different approaches to measuring the
blackbody normalisation in the tail of the bursts. We used an event from
GS~1826$-$24 on 2000 July 1 17:16:37 (\#12 in G08). Four of five PCUs
(0--3) were functioning for this burst. 
We took the approach of iteratively determining the
maximum extent in the tail (beginning from 
a few seconds
after the
burst start) over which a constant fit to the blackbody normalisations was
an acceptable fit to below $3\sigma$ confidence. 
We refer throughout to the best-fit constant value as
$\left<K_{\rm bb}\right>$, to distinguish from the individual $K_{\rm bb}$
values for each time-resolved spectrum.

\begin{figure}[h]
 \vspace{4cm}
 \epsscale{1.2}
 \plotone{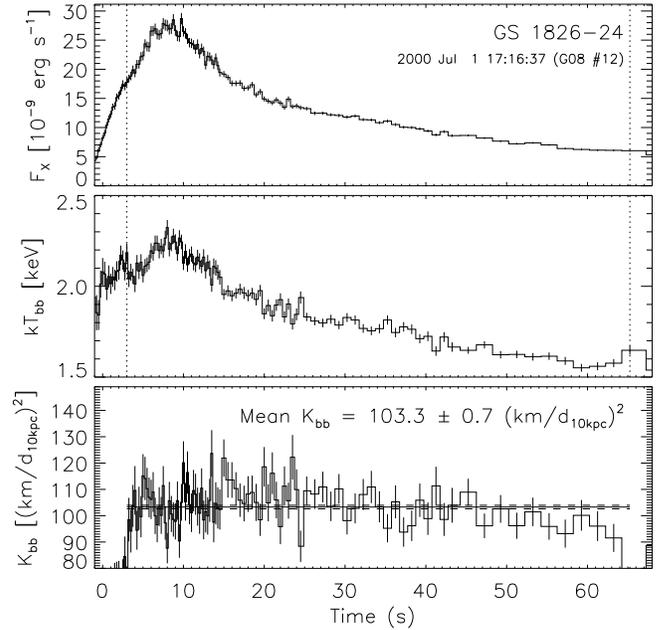}
 \figcaption[]{Example burst observed by \xte\/ from \srca\ on 2000 July  1
17:16:37 UT (\#12 in G08), illustrating the constancy of the blackbody
normalisation over a significant extent of the burst tail. Shown is the
burst flux ({\it top panel}), blackbody (colour) temperature $kT_{\rm bb}$
({\it middle panel}), and blackbody normalisation (in units of $({\rm
km}/d_{\rm 10 kpc})^2$; {\it bottom panel}). The extent over which the
constant-normalisation fit is calculated is illustrated by the vertical
dotted lines ({\it top and middle panels}), and the horizontal lines ({\it
lower panel}), with the $1\sigma$ error range indicated ({\it dashed
lines}). The reduced-$\chi^2=\chi_\nu^2$ for the constant fit is 
$1.45$, for $\nu=95$ degrees of freedom.
 \label{example} }
\end{figure}

We found 
a best-fit 
value of
$\left<K_{\rm bb}\right> =
103.3\pm0.7\ ({\rm km}/d_{\rm 10\,kpc})^2$ over the interval 3.0--65.25~s
(relative to the start over the burst; Fig. \ref{example}).
The reduced $\chi^2/n_{\rm DOF}\equiv\chi^2_\nu$ was 
1.45, for 95 degrees of freedom. 
Remarkably, this consistency was maintained despite the fitted
blackbody temperature $kT_{\rm bb}$ falling from 2.32 to 1.55~keV.

\begin{deluxetable}{lccccc}
 \tabletypesize{\scriptsize}
\tablecaption{Different approaches for determining the mean blackbody
normalisation
 \label{approaches}}
\tablewidth{500pt}
\tablehead{
  & \colhead{Mean $K_{\rm bb}$}
  & \colhead{Time range}
  & \colhead{$kT$ range}
  & \colhead{Flux range}
  & \\
  \colhead{Treatment}
  & \colhead{[$({\rm km}/d_{\rm 10\,kpc})^2$]}
  & \colhead{(s)}
  & \colhead{(keV)}
  & \colhead{[$10^{-9}\, {\rm erg\,cm^{-2}\,s^{-1}}$]}
  & \colhead{$\chi_\nu^2$ (DOF)} 
}
\startdata
averaged, variable bins &
 $103.3\pm0.7$ & 3.0--65.25 & 2.32--1.55 & 28.69--6.09 & 1.45 (95) \\
joint fit, variable binsize &  
 $107.0\pm1.1$ & 11.50--32.25 & 2.16--1.79 & 24.26--11.80 & 1.13 (942) \\
averaged, 0.25-s bins & 
 $103.5\pm0.7$ & 3.25--50.75 & 2.32--1.46 & 28.69--7.05 & 1.28 (186) \\
joint fit, 0.25-s bins & 
 $105.5\pm0.9$ & 4.75--32.00 & 2.32--1.72 & 28.69--11.18 & 1.08 (2460) 
\enddata

\end{deluxetable}

In order to double-check our result, we carried out a similar procedure to
determine the maximum extent over which the blackbody normalisation was
consistent with a constant value, but instead of simply averaging the
fitted normalisations for each time bin, we performed a joint fit in {\sc
XSpec} of the time-resolved spectra to an 
absorbed blackbody model, as
used for the individual spectra. 
We froze the neutral column density at
the same value as used for the individual spectral fits,
$4\times10^{21}\,{\rm cm^{-2}}$. We linked the blackbody
normalisation for each of the individual spectra but allowed
the blackbody temperatures to vary.
With this treatment, the fitted $K_{\rm bb,joint}$ was found 
over a
shorter time interval, between 11.5--32.25~s following the burst start,
and at a 
3\% higher value of $107.0\pm1.1$ (Table \ref{approaches}).
Likely, the higher value is attained by omitting some of the spectra in
the range 40--60~s, when there is a noticeable downward trend in the
normalisation (Fig. \ref{example}). We note that the mean of
the normalisation values over the reduced extent of the burst permitted by
the joint fits, returns a consistent best-fit value, of $106.6\pm1.1$.

\begin{figure}[ht]
\epsscale{1.2}
 \plotone{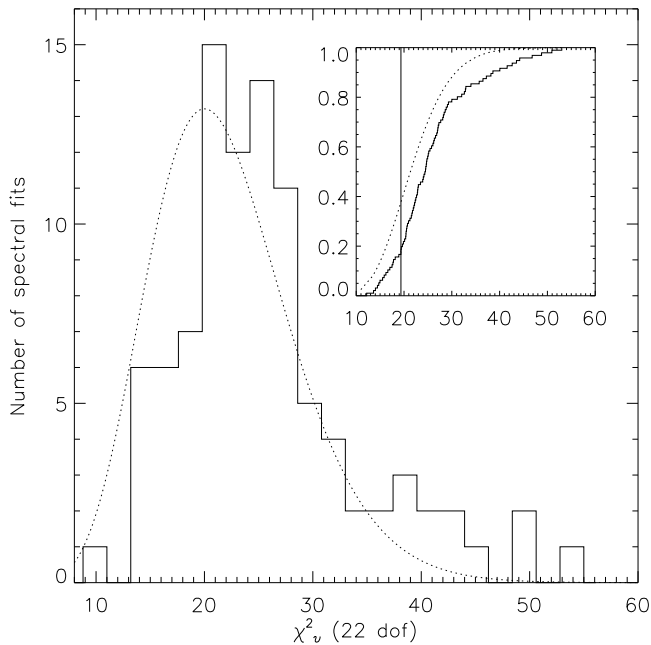}
 \figcaption[]{Distribution of fit statistic $\chi^2$ for blackbody 
fits to time-resolved spectra from burst \#12 from \srca, on 2000 July 1
17:16:37, over the time for which the blackbody normalisation was
determined to be constant 
(3.0--65.25~s relative to the burst start).
The
expected distribution for statistically acceptable fits with 22 degrees of
freedom is overplotted ({\it dotted line}). The fit $\chi^2$  show
an excess of high values, as also indicated by the cumulative probability
distributions ({\it inset}); a K-S test returns a value of 
0.21, indicating disagreement at the $3.4\sigma$ level.
 \label{chidist} }
\end{figure}

The fact that the two treatments
returned a different extent of the burst tail over which the normalisation
was constant may be attributed to the degree of goodness of fit of the
individual spectra. The distribution of $\chi^2$ for the individual fits
in the time range 
3.0--62.5~s
following the burst start is shown in
Fig. \ref{chidist}. Compared to the expected distribution given the number
of degrees of freedom (22), the observed distribution is skewed
significantly to higher values of $\chi^2$.
This result suggests that
residual systematic errors (either instrumental, or possibly from
deviations from a blackbody) prevented a joint fit to the same confidence
level extending over the same extent of the burst tail, compared to the
simple average.

We also explored the effect of the time binning approach on the results,
by repeating the two analyses on data which used uniform 0.25~s bins
throughout the burst. 
The bin sizes for the G08 data were
0.25-s from the start of the burst through to 14.5~s after the start,
0.5~s to 24.5~s, 1~s to 43.25~s, and 2~s through to 65.25~s.
We found that the $\left<K_{\rm bb}\right>$ value obtained with uniform
0.25-s bins
was identical to that with bins of variable size, although the extent 
of the constant fit
was slightly less (extending to 
50.75~s instead of 65.25~s;
Table \ref{approaches}).
Similarly, the joint fit to the uniform 0.25-s binned spectra gave a comparable
result to the joint fit for variable bins, and over roughly the same
extent of the burst.
The $\chi^2$
distribution for the 0.25~s-binned data over the range 
3.25--50.75~s after
the burst start was similarly discrepant from the expected distribution
assuming a statistically good fit, but at a lower confidence level (K-S
statistic of 0.12, equivalent to $2.5\sigma$).

The \xte\/ PCUs are subject to a short ($\approx10\ \mu{\rm s}$)
interval of inactivity following the detection of each X-ray photon. This
``deadtime'' reduces the detected count rate below what is incident on the
detector (by approximately 3\% for an incident rate of
400~count~s$^{-1}\,{\rm PCU}^{-1}$). For \srca, the bursts peak at
$\approx2500$~count~s$^{-1}\,{\rm PCU}^{-1}$, giving a peak deadtime rate
of $\approx6$\% (reducing in the burst tail to
$\approx4$\% after $\approx60$~s). Since the bursts reach approximately the
same peak flux, the deadtime correction is also approximately the same,
and can be neglected for comparison purposes (we similarly neglect any
issues related to the absolute PCA calibration).
However, since the deadtime correction varies with time, it might also be
expected to result in a slightly different evolution of the normalisation
throughout the burst. To test this possibility, we examined the duration
of the consant interval for the burst analysed in this section (\#12 from
G08), with and
without the deatime correction. Although we measured an increase in the
average normalisation with the deadtime correction, as expected, we found
no change in the extent over which the normalisation was found to be
constant. Thus, we neglect deadtime corrections for the remainder of the
analysis in this paper.  

 \subsection{The full burst sample from GS 1826$-$24}
 \label{subsec1}

We analysed 67 bursts observed from GS~1826$-$24 by \xte\/ between 1997
and 2007. This sample includes the set of 58 complete bursts in G08,
excluding seven events for which the full lightcurve was not observed, or
for which no appropriate data modes were available for the analysis, and
one event which occurred during a slew and for which the burst flux could
not be determined correctly\footnote{The excluded bursts are \#8, 15, 21,
29, 39, 42, and 44 from Table 5 of G08}.
We also analysed an additional 9 bursts from observations in 2006 August,
which were not part of the G08 sample.

We compared the results of the joint fits $K_{\rm bb,joint}$ with the
average of the individual fits $\left<K_{\rm bb}\right>$ for the bursts
from this sample to determine how the results from our
sensitivity tests 
translated to a larger sample. Although in some cases the joint fit
durations were shorter than the constant fit duration for the individual 
normalisations
the median duration was 94\% of the average interval.
Furthermore, we found no systematic difference between the normalisations
determined by the two methods, and the agreement was at better than 2\%
for 76\% of the measurements (maximum deviation was 8.5\%). This compares
favourably to the typical $1\sigma$ statistical uncertainty on the
measurements, of 1\%. Thus, we conclude that the parameter averages
$\left<k_{\rm bb}\right>$
provide an unbiased measure of the blackbody normalisation in the tail of
these bursts, and adopt that method for further analysis.

\begin{figure}
\epsscale{1.2}
 \plotone{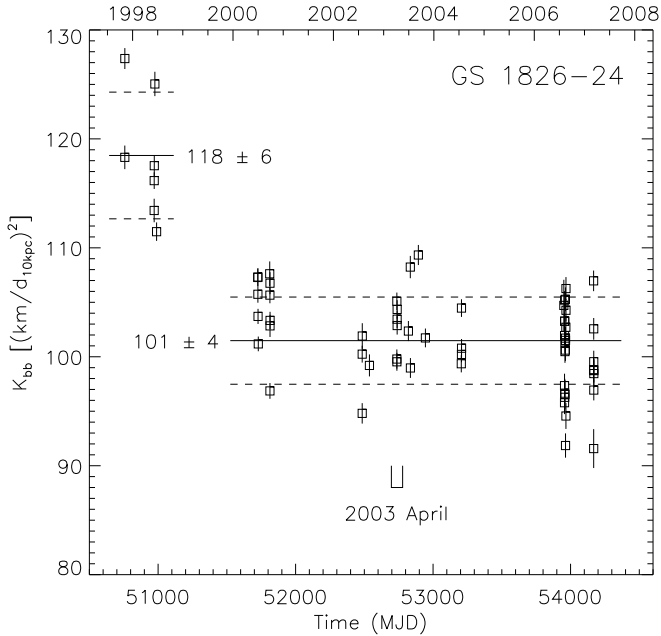}
 \figcaption[]{Mean blackbody normalisation values for 67 thermonuclear
bursts observed by \xte\/ from \srca. The mean and standard deviation for
the two observation epochs (1997--8 and 2000--7) are indicated.
Note the marked discrepancy between the mean values for the two
distributions; the K-S statistic 
indicate that they are discrepant at a significance of $1.3\times10^{-6}$,
equivalent to $4.7\sigma$.
 \label{history} }
\end{figure}

The 
$\left<K_{\rm bb}\right>$ for the bursts from \srca\ determined 
from the individual spectral fit parameters 
(using the spectra with variable binsizes; see \S\ref{sec1}) is shown in
Fig. \ref{history}. The typical span of the constant fit was from 
$<3$~s after the burst start (52\% of the bursts) 
to approximately 60~s. For one burst
the normalisation was found to be constant from 
2--77~s
after the start of
the burst; the median duration was 
48~s.
Over the interval during which the blackbody normalisation was constant,
the blackbody temperature typically decreased from 2.3 to 1.6~keV, 
while the flux decreased from $\approx3\times10^{-8}\ \epcs$ to
$\approx5\times10^{-9}\ \epcs$. 

The $\left<K_{\rm bb}\right>$ values were not significantly correlated
with either the duration over which the average was calculated, 
nor the maximum time out
to which the average was calculated.
Interestingly, the earliest time to which the constant fits could be
extended ($\approx1$--3~s after the burst start) was within the
$\approx5$~s burst rise. For several of the bursts, the time at which the
normalisation reaches the mean value corresponds approximately
to a change in slope of the burst rise, seen in several of the bursts
(e.g. Fig. \ref{example}). This feature is similar to the ``bump'' seen in
model lightcurves \cite[e.g.][]{heger07b}. Although in the simulations
this feature must arise due to some variation in the rate of increase of
thermonuclear energy production
(as the 1-D models cannot account for lateral propagation),
the coincidence of the observed change in
slope with the normalisation achieving its mean value suggests that this
point marks where the effects of spreading cease. Further analysis of the
detailed shape of the burst rises may help to clarify this situation.

We found significant variations in the mean normalisation
$\left<K_{\rm bb}\right>$, both between
and within different observational epochs.  
For the bursts within each epoch (1997--8 and
2000--7), the normalisation varied by 5 and 4\%,
respectively.  This variation was highly significant; the $\chi^2$ values
for constant fits were 218 (for 6 degrees of freedom) and 
1007 (for 59
degrees of freedom), respectively.
In terms of the inferred radius, this implies systematic uncertainties of
order 2--2.5\%. 
Additionally, a much larger variation was measured between the two epochs.
The first seven bursts, observed between 1997 November and 1998 June, had
$\left<K_{\rm bb}\right>$ values substantially larger than the remaining
bursts (observed from 2000 June onwards). The
mean and standard deviation for the normalisations of the 1997-8 bursts was 
$118\pm6\ ({\rm km}/d_{\rm 10\,kpc})^2$
while for
subsequent bursts was 
$101\pm4\ ({\rm km}/d_{\rm 10\,kpc})^2$. A two-sided Kolmogorov-Smirnov
test confirms that these distributions are discrepant at the $1.3\times10^{-6}$
significance level (equivalent to $4.7\sigma$).
This variation was also noted by \cite{gal03d}, who
reported instead a correlation between the
persistent flux and the blackbody normalisation for a subset of the
bursts analysed here.

\begin{figure}
\epsscale{1.2}
 \plotone{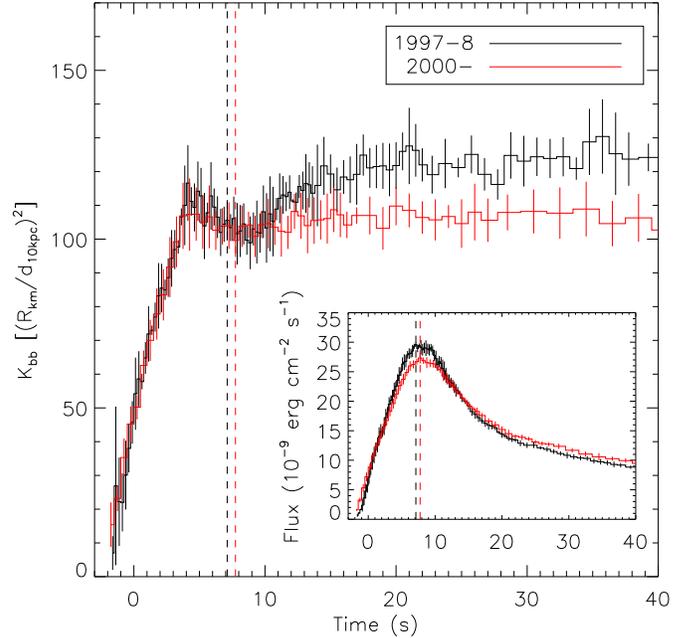}
 \figcaption[]{Comparison of averaged blackbody normalisation profiles for
bursts from \srca\ measured in 1997--8 (\#1--5 of G08) and 2000--7 (\#9,
10, 11, 12, 13, 16, 17, 19, 20). The vertical dashed lines indicate the
time of maximum flux for each set of bursts. Note the agreement in the
normalisation throughout the burst rise and maximum, and the increasing
discrepancy from 10~s after the burst start. 
The inset shows the corresponding variation of the averaged burst flux.
\label{epoch} }
\end{figure}

The observed variation is unlikely to arise from any instrumental effect,
as the PCA is precisely calibrated to maintain stable flux measurements
for calibration sources over the entire mission lifetime. As we discuss
below, the discrepancy is also unlikely to result from variations in the
neutral column density $n_H$ as a function of epoch. Closer examination of
the spectral variation in the two groups of bursts provides a possible
explanation. In contrast with the example burst discussed in the previous
section, and other bursts observed in 2000--7, the constant fit interval
for the bursts observed in 1997--8
began later than in the 2000--7 bursts; typically 10~s after the burst
start for the 1997--8 bursts, compared to 3~s after the burst start for
the other bursts. To put this another way, there was additional variation
in the
blackbody normalisation during the burst rise for the 1997--8 bursts that
prevented extension of the constant interval to the same point as in the
later bursts. This is illustrated in Fig. \ref{epoch}, which compares the
blackbody normalisation averaged over the 1997--8 bursts with that of the
2000--7 bursts. Remarkably, however, the behaviour of the normalisation in
the two groups of bursts is virtually identical during the burst rise; the
discrepancy sets in from  10~s after the burst start, with the
normalisation of the 1997--8 bursts gradually increasing over another
$\approx10$~s to a higher level.

 \subsection{KS~1731$-$26}
 \label{s1731}

\xte\/ observations of \srcb\ in 2000 August--September detected 14
bursts, 8 in August and 6 in September 
(these are bursts \#14--21 and 22--27 in G08, respectively). 
Although the \xte/PCA observations 
were interrupted regularly
due to the satellite orbit and other scheduled observations, the times of
the detected bursts were consistent with a regular recurrence time.
The shortest burst separation measured by \xte\/ during each month of data
was $\approx2.5$~hr; the longer separations were consistent with integer
multiples of this
value, indicating regular bursts where intervening events were missed
within data gaps. We measured the average recurrence times based on linear
fits to the burst arrival times measured by \xte, as
$2.577\pm0.011$~hr and $2.636\pm0.003$~hr for August
and September, respectively.
One further burst, observed on 1999 Aug 26 (\#13 in G08), exhibited a
lightcurve consistent with the 14 observed in 2000, and we included it in
this sample.

\begin{figure}
\epsscale{1.2}
 \plotone{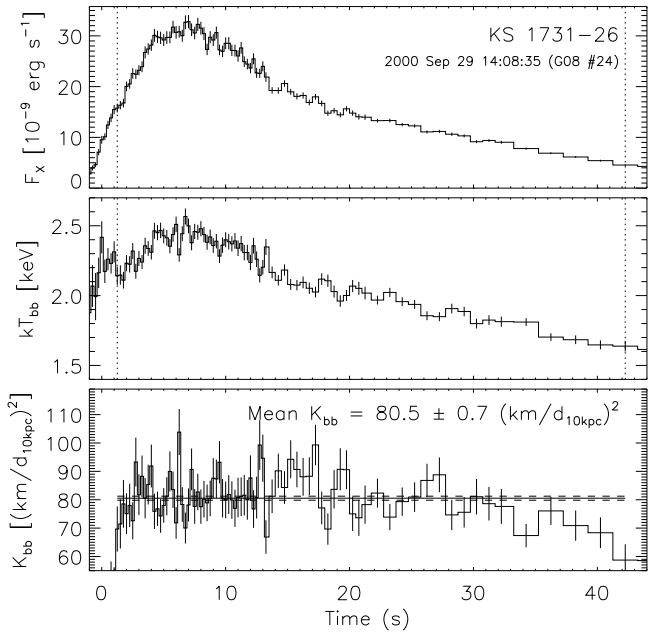}
 \figcaption[]{Example burst observed by \xte\/ from \srcb\ on 
2000 September 29 14:08:35 UT
(\#24 in G08), illustrating the constancy of the blackbody
normalisation over a significant extent of the burst rise and tail. 
Panel descriptions are as for Fig. \ref{example}.
The reduced-$\chi^2=\chi_\nu^2$ for the constant fit is $1.39$, for
$\nu=79$ degrees of freedom.
 \label{lightcurves} }
\end{figure}

The regular bursts featured similar long ($\approx5$~s) rises and decays
($\approx60$~s) as those typically observed from GS~1826$-$24 (Fig.
\ref{lightcurves}). 
The peak count rate was $\approx2300$~count~s$^{-1}$~PCU$^{-1}$, so that
deadtime corrections are comparable to those of \srca\ (see \S\ref{sec1}).
However, the recurrence time for the bursts from
KS~1731$-$26, at $\approx2.6$~hr, were significantly shorter than has been
observed from GS~1826$-$24 in years of \xte\/ observations
\cite[e.g.][]{thompson08}.
Based on the assumed distances for the two sources (6~kpc for GS~1826$-$24
and 7.2~kpc for KS~1731$-$26; see 
G08), the bursts
from KS~1731$-$26 reached significantly higher luminosities of
(1.8--$1.9)\times10^{38}\ \eps$,
and the burst
durations were also significantly shorter ($\tau=24$~s compared with
$\approx40$~s for \srca). A
shorter burst duration for \srcb\ suggests a smaller fraction of hydrogen
in the burst fuel, although the shorter recurrence time also allows less
hydrogen to be consumed by steady burning prior to the burst. 

As with \srca,
we measured the best-fit normalisation $\left<K_{\rm bb}\right>$ 
from the time-resolved blackbody spectral fits
over the longest possible time interval without exceeding the $3\sigma$
confidence limit.
With the shorter bursts from \srcb, the constant $K_{\rm bb}$ fits
extended
typically over the range 
2--30~s after the burst start. Over this
time interval, the blackbody temperature dropped (typically) from 2.5
to 1.8~keV, with the flux dropping 
from $3\times10^{-8}\ \epcs$ to $7\times10^{-9}\ \epcs$.

The fitted $\left<K_{\rm bb}\right>$ values of the long bursts from \srcb\
exhibited significant variability; a fit with a constant model gave a 
$\chi^2=112.7$ for 14 degrees of freedom, indicating variability at the
$8.5\sigma$ level.
The mean value was 
$80\pm2\ ({\rm km}/d_{\rm 10\,kpc})^2$, 
with standard deviation
between the various measurements of 2.6\%. 
The mean blackbody normalisation of the burst on 1999 Aug 26,
measured between 1.25--36~s after the burst start at
$79.7\pm0.8$~(km/$d_{\rm 10\,kpc})^2$, was fully consistent with the
bursts observed in 2000 despite being observed a year earlier.

\begin{figure}
\epsscale{1.2}
 \plotone{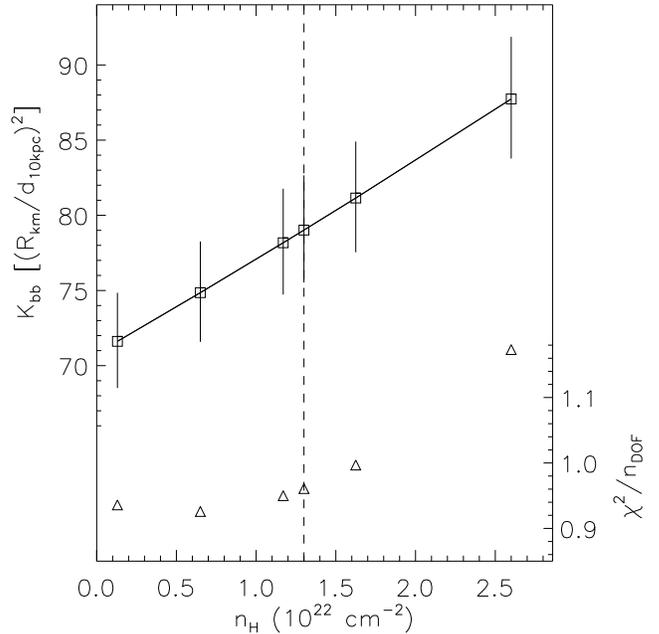}
 \figcaption[]{Sensitivity of the blackbody normalisation in spectral fits
of \xte\/ data to the assumed column density $n_H$. The fits were
calculated from a simulated spectrum with $kT=2.1$~keV, including
pre-burst persistent and instrumental background from burst \#15 of
KS~1731$-$26. The square symbols
show the fitted value of the blackbody normalisation (left-hand $y$-axis)
as a function of the assumed column density $n_H$. 
The lower half of the plot shows the corresponding $\chi^2$ values; each
fit is statistically acceptable.
The choice of the $n_H$ value can clearly
introduce a systematic bias to the normalisation values, of the order a
few per cent.
\label{nH_sens} }
\end{figure}

In an independent analysis of the bursts observed from \srcb\ with \xte\/
(of which the bursts we analyse here are a subset),
\cite{guver11a} measured a mean blackbody normalisation of 
$88.4\pm5.1\ ({\rm km}/d_{\rm 10 kpc})^2$, and found weakly significant
variation in the normalisation as a function of the burst flux. Although
this value is not significantly different from the normalisation we derive
from the subset of bursts analysed here,
we suggest that the difference arises from a systematic effect due to the
value
adopted for the neutral column density.
For the analysis presented here, we assumed 
$n_H=1.3\times10^{22}\ {\rm cm^{-2}}$ (from multi-epoch {\it Chandra}\/ and
{\it XMM-Newton}\/ spectra of the source during quiescence;
\citealt{cackett06}) while \cite{guver11a} adopted a value of
$2.98\times10^{22}\ {\rm cm^{-2}}$, the mean of best-fit values from fits
to the burst spectra measured by {\it RXTE}.
For a simulated blackbody spectrum
with known $kT$ and normalisation,
adopting a realistic model for the pre-burst emission as background, the
fitted value of the normalisation depends linearly on the assumed $n_H$
(Fig. \ref{nH_sens}). 
That is, over (under) estimating the $n_H$ value assumed for the fits will
have the effect of over (under) estimating the $K_{\rm bb}$.
For the representative parameters chosen, the slope gives
approximately $6.5\ ({\rm km}/d_{\rm 10 kpc})^2$ for every additonal
$10^{22}\ {\rm cm^{-2}}$ in column that is adopted. The difference
between the adopted values for the two analyses could account for an
offset in the normalisations of up to $11\ ({\rm km}/d_{\rm 10 kpc})^2$,
sufficient to explain the discrepancy.

\section{Discussion}
\label{sec2}

We found significant variations in the blackbody normalisation 
 $\left<K_{\rm bb}\right>$ averaged
over tens of seconds of the burst tail
in a homogeneous sample of regular, consistent bursts from \srca\ and
\srcb,
both within and between observation epochs. 
We found that the
variation within epochs was 4--5\% (2.6\%) for \srca\ (\srcb),
while the variation between epochs can be as large as 17\% (for \srca).
In terms of
inferred radius, this corresponds to variations of 2--2.5\%
(1.3\%) and 8\%.
The variation within each epoch is within the range reported by
\cite{guver11a} in their more comprehensive study of bursts from several
sources, although the 8\% variation between epochs observed for \srca\ is
somewhat larger.
We stress that this uncertainty is solely related to the measurement of
the blackbody normalisation in bursts, and (as suggested by Z11) may be
exceeded by other sources of uncertainty, for example the degree of
anisotropy of burst emission.

The
1997--8 bursts from \srca, which exhibited longer recurrence times and
reached slightly higher fluxes, also exhibited larger mean normalisations
$\left<K_{\rm bb}\right>$.
Comparison of the normalisation time-series averaged over
subsamples of bursts from the two epochs 
show
that the normalisations were essentially identical during the burst rise
and through the burst peak, with deviations becoming apparent between
10--20~s after the burst start.  The reproducibility of the $K_{\rm bb}$
evolution throughout the burst rise and peak suggests that the system
geometry, atmospheric composition and temperature (and hence $f_c$) were
essentially identical
over that interval,
and whatever physical condition gave rise to the
elevated $K_{\rm bb}$ in the 1997--8 bursts set in after the burst peak.
We consider three possible mechanisms to give rise to the variation.

First, it may be that the amount of neutral material close to the neutron
star was reduced, so that the neutral column density $n_H$ decreased
during the 1997--8 bursts, leading to an overestimation of $K_{\rm bb}$
(see e.g. Fig. \ref{nH_sens}). Simulations adopting the pre-burst emission
from \srca\ (as was done for \srcb; see \S\ref{s1731}) indicate that in
order to overestimate the $\left<K_{\rm bb}\right>$ by the required amount
would necessitate decreasing the total $n_H$ value by 
$1.9\times10^{22}\ {\rm cm^{-2}}$. Since this value is almost five times
the assumed line-of-sight value for \srca, we can rule out this mechanism.

Second, it is possible that the degree of anisotropy of the burst emission
changed in reponse to a variation in the accretion geometry
\cite[e.g.][]{fuji88}, perhaps triggered by the burst.
Such a change might be expected to
be reflected in the X-ray spectrum. The broadband ($\approx1$--100~keV)
spectral distribution of the persistent emission in \srca\ has been well
studied by \cite{thompson08}, in the context of determining the best
possible estimate of the bolometric flux (and hence the accretion rate).
Those authors found substantial changes in the spectrum with epoch, most
notably during 2003 April, when evidence for an additional soft ($<1$~keV)
component was found. However, the mean normalisation from the 6 bursts
observed in 2003 April was $(102\pm4)\ ({\rm km}/d_{\rm 10 kpc})^2$ (see Fig. \ref{history}), fully
consistent with the mean from the other bursts from 2000 onwards (but
excluding the 2003 April bursts), of $(101\pm4)\ ({\rm km}/d_{\rm 10
kpc})^2$.
Conversely, there is no evidence for a substantially different X-ray
spectrum during the 1997--8 observations (see \citealt{thompson08}, Fig.
4)
The  lack of correspondence
between the broadband spectral shape and the blackbody normalisation from the
bursts seems to contraindicate an influence of the burst anisotropy on the
normalisation, although cannot rule it out.
On the other hand, the variations in the persistent X-ray spectrum might be
transient, present only when the elevated $K_{\rm bb}$ values were
measured, beginning 10~s
after the burst start. 
Such variations might be expected to lead to incorrect pre-burst emission
subtraction later in the burst, although this is not observed
consistently.

Third, we consider the possibility that the $f_c$ value changes to produce
the variation in the measured $K_{\rm bb}$ after the peak in the 1997--8
bursts from \srca. Modelling studies indicate that $f_c$ depends on fixed
parameters such as the neutron star surface gravity, but also the
composition and effective temperature of the scattering atmosphere
\cite[e.g.][]{madej04,suleimanov11a}. There is evidence in other sources that
radius-expansion bursts can remove the outer, H-rich layers of the
photosphere, leading to a change in the atmospheric composition during the
burst \cite[]{gal06a}; no such effects have been suggested from
non-radius expansion bursts. Nevertheless,
interpreting the maximum variation in $\left<K_{\rm bb}\right>$ in the
1997--8
bursts, compared to the mean for the 2000--7 bursts, as a variation in
$f_c$ implies a maximum variation of 12\% (8\% in the mean).
This result suggests that the assumption that $f_c$ is constant
during bursts is not always true.
While \cite{suleimanov11a} predict patterns of variation
of $f_c$ with burst flux, the epoch-to-epoch differences in the
burst lightcurves in \srca\ are relatively subtle, and do not appear
sufficient to give rise to the inferred variation in $f_c$.

It is possible that
the peak blackbody flux (or temperature) serves as the discriminant which
results in elevated blackbody normalisations later in the burst.
The 1997--8 bursts reached
maximum fluxes about 8\% higher on average than for the 2000--7 bursts.
The discrepancy in the maximum temperature reached was proportionately
smaller.
The samples of bursts studied here were deliberately selected for the
consistency of their lightcurves and regularity of their recurrence times.
For other samples of bursts, which are typically much
more heterogeneous,
systematic errors in
measurements of the blackbody radius of order $\gtrsim8$\% 
should be assumed, unless the variation in $f_c$ can be modelled.

The mean blackbody normalisations measured during the
later (2000--7) bursts from \srca, and the regular bursts from \srcb, 
were consistent with a constant value over several tens of
seconds. This 
constancy was observed independent of the specific method for
determining the mean value, although the choice of method can introduce small
biases. In particular, comparisons of joint fits to
the spectra with averages of the fitted normalisations show that in
general the former method arrives at shorter durations for the constant
normalisation, likely because the spectra are not (en masse) statistically
consistent with blackbodies.
Trends in the blackbody normalisation late in the burst tail, coupled with
these marginally deviant spectra, likely give rise to systematic
errors of a few percent 
between the two methods.
The choice of time binning strategy does not have as large an effect.

This constancy of the blackbody normalisation was maintained
despite significant decreases in the blackbody temperature, of (typically)
30\%, and decreases in the flux of 60--75\% over the same time interval.
Such a lack of variation in the blackbody normalisation
implies constraints on the relative degree of spectral distortion over
this temperature range, 
suggesting one of two situations. Either, any variation in in the spectral
distortion factor $f_c$ as a
result of the varying effective temperature (as indicated by the
decreasing blackbody temperature) is {\it exactly balanced}\/ by some
other variation, e.g. a change in
the emitting radius as the burst flux decreases; or, that the
color temperature correction $f_c$ is also constant throughout the 
interval in which $K_{\rm bb}$ is constant.
The former explanation is rather contrived, particularly considering that
in the burst tail the burning front is expected to have already spread to
the entire surface area of the neutron star, and no further increase (or
decrease) in burning area is expected. Thus, these measurements suggest
that 
for the range of effective temperatures spanned in the
burst tails, the color-temperature correction $f_c$ is roughly constant.
This conclusion is difficult to reconcile with the predictions of
atmosphere models \cite[e.g.][]{suleimanov11a}, which indicate significant
variations in $f_c$ over most of the flux range spanned during the burst.
The distance for \srca\ is thought to be $\approx6$~kpc \cite[]{heger07b},
at which an Eddington-limited burst would be expected to reach
$F/F_{\rm Edd}=3.7\times10^{-8}$ ($6.3\times10^{-8}$) $\epcs$ for an
H-rich (pure He) atmosphere (e.g. G08). This would imply that the range of
flux over which the burst
normalisation is found to be constant is 0.16--0.75
(0.10--0.46)~$F/F_{\rm Edd}$. 
The greatest variation in the normalisation predicted by
\cite{suleimanov11b} is outside these ranges, although significant
variations would yet be expected (particularly for the solar metallicity
models; see also Z11).

The $\approx3$--5\% fractional variation in $\left<K_{\rm
bb}\right>$ observed
from \srca\ and \srcb\ within each observational epoch
was smaller than that seen in two
radius-expansion bursts observed from EXO~1745$-$248, of 25\%
\cite[]{ozel09a}. This observation perhaps 
suggests an explanation 
of the variation. EXO~1745$-$248 exhibited during it's 2000
outburst an initial period of strong variability, reminiscent of dipping
behaviour observed in high-inclination systems \cite[e.g.][]{bcatalog}. No
such variability has ever been observed from \srca\ or \srcb, suggesting
that perhaps the inclination is lower in those systems than in
EXO~1745$-$248. 
If system inclination is the main factor in determining the variation in
apparent blackbody normalisation, 
a possible explanation is the reprocessing of some fraction of the
burst flux off an accretion disk whose projected area varies with time. In
4U~1728$-$34, the timescale inferred for the variation was several tens of
days \cite[]{gal03b}, much longer than the expected orbital period of this
system \cite[e.g.][]{gal10b}. However, in \srca\ and \srcb, the variation
is on a much shorter timescale, comparable to the recurrence time of the
bursts themselves (hours). For comparison, the discrepant bursts from
EXO~1745$-$248 were separated by 8.5~d.
We note that such an explanation fails to account for the
factor of $\approx2$ difference in normalisation in the bursts from
4U~1724$-$307 \cite[]{suleimanov11b}, which does not show dips and thus is
unlikely to be at high inclination. However, those bursts also exhibited
markedly different timescales, indicative of a varying accumulated fuel
reservoir at ignition; we suggest instead that different physical
conditions (temperature, composition) gave rise to the difference in the
measured blackbody normalisation. 

The shorter timescale of variation
in 
$\left<K_{\rm bb}\right>$ for 
\srca\ and \srcb\ suggests that
there may be an orbital component of the variation. The specific value of
the blackbody normalisation depends upon the assumed neutral column
density, as illustrated in Fig. \ref{nH_sens}. Thus, an orbital variation
in the line-of-sight column density, perhaps arising from cool clouds of
material above the point of contact of the accretion stream with the disk,
will manifest as a variation in the blackbody normalisation on the same
timescale.
In order to test this hypothesis, we calculated a Lomb-normalised
periodogram on the blackbody normalisation measurements as a function of
time. 
Recall the orbital period in both sources is unknown, but for \srca\ is
thought to be around 2~hr; based on the typical orbital periods for other
burst sources, we searched a frequency range of 0.5--24~hr.
For \srca, we divided the measurements from the two epochs through
by the appropriate mean, and found a peak Lomb power of 10.48;
for
\srcb, we found a peak Lomb power of 5.29. Neither of these detections is
significant.
For \srcb, the 
known sensitivity of the normalisation measurement to discrepancies
between the assumed and true value of $n_H$ indicates that a variation of 
$0.3\times10^{22}\ {\rm cm^{-2}}$ in $n_H$ could account for the variation
in the measured blackbody normalisation. For \srca, the measurements are
slightly more sensitive to discrepancies in $n_H$, so that a variation in
$n_H$ of $0.45\times10^{22}\ {\rm cm^{-2}}$ could account for the
variation in the blackbody normalisation within each epoch.
If orbital or longer-timescale variations in $n_H$ are driving the
variation in the blackbody normalisation, it may be possible to verify
through time-resolved high-spectral resolution measurements of absorption
edges in the X-ray spectra.

\acknowledgments

We thank the anonymous referee, who made several substantial comments
which improved the paper.
DKG is a member of an International Team in Space Science on 
type-I X-ray bursts sponsored by the International Space Science Institute
(ISSI) in Bern, Switzerland, and we thank ISSI for hospitality during part
of this work.
This research has made use of data obtained through the High Energy
Astrophysics Science Archive Research Center Online Service, provided by
the NASA/Goddard Space Flight Center.  
DKG is the recipient of an Australian Research Council Future Fellowship
(project FT0991598).

Facilities: \facility{RXTE}

\end{document}